\documentclass{PoS}
\usepackage{subfig}
\usepackage{url}
\usepackage{array}
\usepackage{makecell}

\usepackage{lineno}

\title{H.E.S.S. searches for TeV gamma rays associated to high-energy neutrinos}
\ShortTitle{H.E.S.S. neutrino follow-up program}

\author{\speaker{Fabian Sch\"ussler}\\
        IRFU, CEA, Universit\'e Paris-Saclay, F-91191 Gif-sur-Yvette, France\\
        E-mail: \email{fabian.schussler@cea.fr}}
\author{H.~Ashkar, M.~Backes, K.~Egberts, F.~Brun, M.~F\"u{\ss}ling, C.~Hoischen, J.-P.~Lenain, I.~Lypova, S.~Ohm, D.~Parsons, C.~Romoli, M.~Seglar-Arroyo, M.~Zacharias, A.~Zech on behalf of the H.E.S.S. Collaboration\footnote{for collaboration list see PoS(ICRC2019)1177}}

\abstract{
The detection of an astrophysical flux of high-energy neutrinos by IceCube is a major step forward in the search for the origin of cosmic rays, as this emission is expected to originate in hadronic interactions taking place in or near cosmic-ray accelerators. No neutrino point sources, or significant correlation with known astrophysical objects, have been identified in the IceCube data so far. The hadronic interactions responsible for the neutrino emission should also lead to the production of high-energy gamma rays. The search for neutrino sources can then be performed by studying the spatial and temporal correlations between neutrino events and very high energy (VHE, E > 100 GeV) gamma rays. 
We report here on the search for VHE gamma-ray emission with the H.E.S.S. imaging air Cherenkov telescopes (IACTs) at the reconstructed position of muon neutrino events detected by IceCube. We will provide an up-to-date summary of the extensive program to perform prompt IACT observations of realtime IceCube neutrino event positions. A recent highlight of this program are the H.E.S.S. observations during the broad multi-wavelength campaign that followed the detection of the neutrino event IceCube-170922A arriving from a direction consistent with the location of a flaring gamma-ray blazar TXS 0506+056 in September 2017. We'll present the H.E.S.S. observations obtained within ~4h hours of the neutrino detection as well as a complementary search for gamma-ray emission at longer timescales and put them into the multi-wavelength and multi-messenger context. 
}

\FullConference{36th International Cosmic Ray Conference -ICRC2019-\\
		July 24th - August 1st, 2019\\
		Madison, WI, U.S.A.}

\begin{document}

\section{Introduction}
Six years ago the IceCube collaboration announced the detection of a diffuse flux of astrophysical neutrinos~\cite{HESE1} thus opening the window to high-energy neutrino astronomy. So far the underlying sources have not been identified. The recorded neutrino arrival directions are consistent with an isotropic flux (modulated by the IceCube acceptance) and no significant correlations with the Galactic Plane has been established. An extragalactic origin is therefore favoured. The astrophysical flux is significant at energies between $\sim20$ TeV and a few PeV and its energy spectrum is consistent with a $E^{-\Gamma}$ power law, with IceCube analyses reporting spectral indices $\Gamma$ in the 2.1--2.5 range~\cite{Leif}.  

A promising way to search for the sources of astrophysical neutrinos is the multi-messenger approach which combines observations in different wavelength and astrophysical messengers like neutrinos. Here we focus on the combination of observations of neutrinos and very high-energy gamma-rays. Neutrinos are expected to be produced in the decay of pions originating in cosmic-ray interactions and the same process should give rise to a flux of hadronic gamma rays. If they are not absorbed by interactions in the supposedly dense regions at the source or during the propagation within the extragalactic background radiation fields, these gamma rays could be detected at Earth. 

\begin{figure}[!t]
\centering
\includegraphics[width=1.\textwidth]{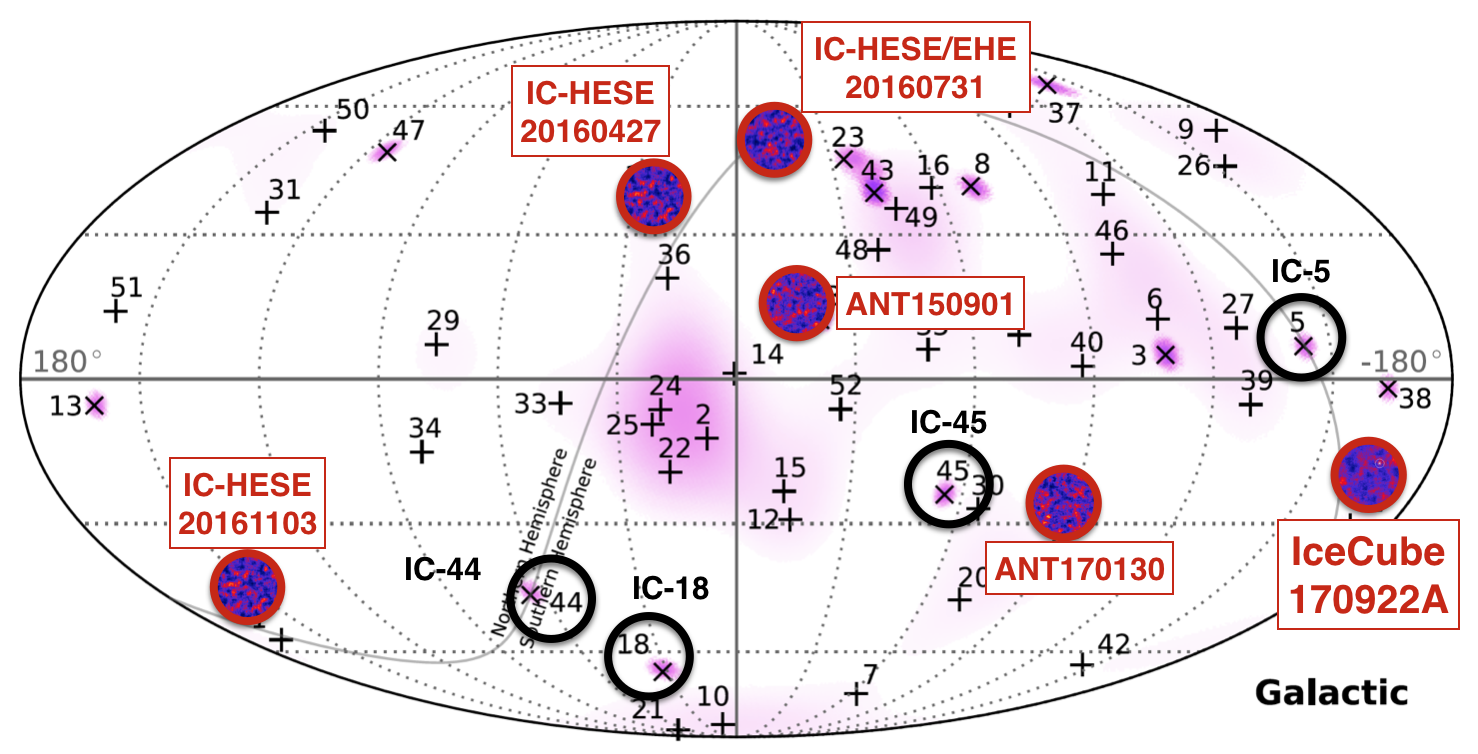}
   \caption{Overview of the H.E.S.S. neutrino follow-up program showing the high-energy neutrino sky represented by events detected by IceCube. Archival H.E.S.S. observations around the events IC-5, IC-18, IC-44 and IC-45 are highlighted by black circles. Follow-up observations of neutrino alerts from the IceCube and ANTARES neutrino telescopes have been performed around the regions denoted in red. Modified from~\cite{HESSMM_ICRC2017}.}
\label{fig:overview}
 \end{figure}
 
As part of its extensive multi-messenger program~\cite{HESSMM_ICRC2017}, the H.E.S.S. collaboration has been actively searching for gamma ray emission associated to high-energy neutrinos. Here we present a status update on these searches which are illustrated in Fig.~\ref{fig:overview}. We specifically focus on a recent highlight of the program: the follow-up campaigns on TXS~0506+056 and the recent inclusion of H.E.S.S. to the Gamma-ray Follow-Up (GFU) program of the IceCube collaboration.

\section{IceCube-170922A and TXS~0506+056}
\begin{figure}[!t]
\centering
\includegraphics[width=1.\textwidth]{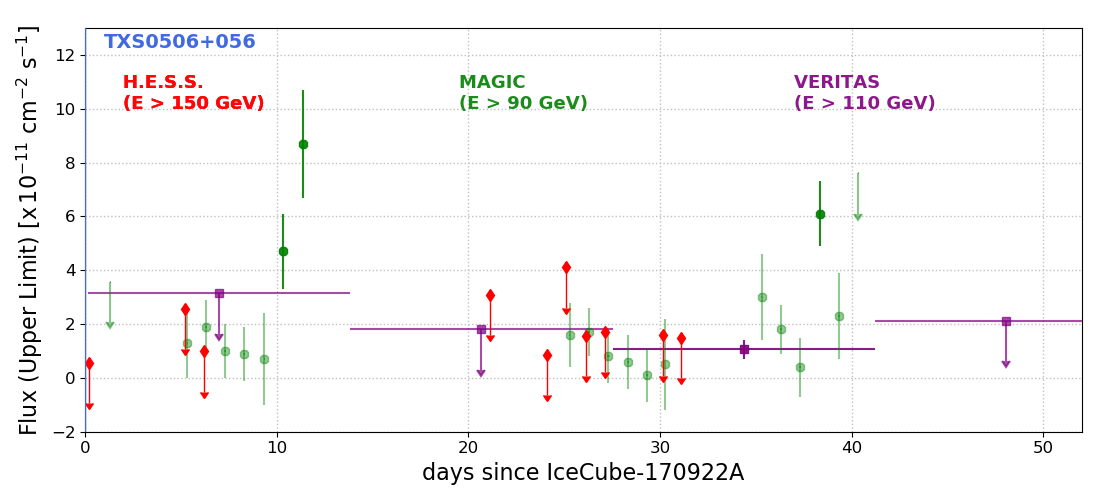}
   \caption{Lightcurve of VHE gamma-ray observations of the blazar TXS 0506+056 obtained following the alert on the detection of the high-energy neutrino IceCube-170922A. Red markers denote the upper limits derived from H.E.S.S. observations, while the green and magenta markers show the data obtained by the MAGIC~\cite{MAGIC170922} and VERITAS~\cite{VERITAS170922} IACTs.}
\label{fig:IACTs170922}
 \end{figure}

On September 22nd, 2017 IceCube detected an interesting high-energy neutrino event (EHE 170922 or IceCube-170922A). An alert was released by the IceCube real-time system within 43 seconds, triggering follow-up observations across several MWL bands. First among the VHE observatories, H.E.S.S. observations started 4 hours later as soon as the region became visible and continued during the next night. No significant gamma-ray emission was detected~\cite{ATEL:HESS170922}. Several days later it was realized that the reconstructed direction of the neutrino event was only 0.1$^{\circ}$ away and fully compatible with the sky position of the BL Lac object TXS~0506+056, which had shown historically high level of activity in all wavelengths, most notably in GeV gamma rays monitored by {\it Fermi}-LAT~\cite{ATel:Fermi170922} over several months prior to the neutrino alert. Renewed and deeper VHE follow-up observations were obtained by all IACT collaborations over the following days, weeks and months. The source was detected for the first time by the MAGIC telescopes~\cite{ATel:MAGIC170922, MAGIC170922}, later followed by the VERITAS observatory~\cite{VERITAS170922}. Due to the extremely rapid variability of the source at the highest energies and unfavorable observing conditions no VHE gamma-rays have been detected by H.E.S.S. A summary of the observations is given in Fig.~\ref{fig:IACTs170922}.

The chance probability for a high-energy neutrino to be detected in coincidence with a flaring blazar from the {\it Fermi}-LAT catalogues was found to be disfavored at the 3$\sigma$ confidence level~\cite{IC-TXS-MMA}. The significance of the TXS~0506+056 and IceCube-170922A association is therefore too low to claim the discovery of the first neutrino source. Nevertheless, the event sparked significant interest in the broader community and, thanks to the extensive MWL data, provides a unique opportunity to study the interplay between energetic photons, neutrinos and cosmic rays. Moreover, a search for previous neutrino emission from the identified direction revealed a neutrino flare of several months duration in 2014-2015 in the IceCube data. This flare, which seems not correlated with any increased high-energy electromagnetic emission in Fermi-LAT has a significance of about $3.5\sigma$. Currently undisputed theoretical models trying to explain both the correlation of IceCube-170922A and the gamma-ray flare of TXS~0506+056 as well as the {\it orphan neutrino flare} are still missing (e.g.~\cite{MMpictureTXS}). Further joint observations of this kind will allow to shed light on the potential association between high-energy neutrinos and flaring blazars and may provide the long-sought hints for the sources of cosmic rays. While the neutrino follow-up programs continue with the current instruments like H.E.S.S., MAGIC and VERITAS, this first exciting result has already strong influence on the preparations of next-generation observatories like the Cherenkov Telescope Array~\cite{CTATransients_ICRC2019, CTANeutrinos_ICRC2019}.

\subsection{H.E.S.S observations of TXS~0506+056: follow-up of IceCube-170922A}
The H.E.S.S. follow-up observations of the neutrino event IceCube-170922A started as soon as the direction became visible in the nights of Sept 22nd \& 23rd. The first observation started $\sim$4 hours after the circulation of the neutrino alert~\cite{icecubegcn}. A preliminary on-site analysis did not reveal any significant gamma-ray emission~\cite{HessATEL} for this data set. A second set of observations was acquired during the nights of Sept 27th \& 28th following the announcement of {\it Fermi}-LAT that TXS~0506+056 was in an active state~\cite{fermiATEL}. The analysis of this early data set of a total of 3.25 hours did not reveal significant gamma-ray emission~\cite{IC-TXS-MMA}.

\begin{figure}[!t]
  \centering
   \subfloat[excess map]{\label{fig:TXSearly:excess}
    \includegraphics[width=0.48\textwidth]{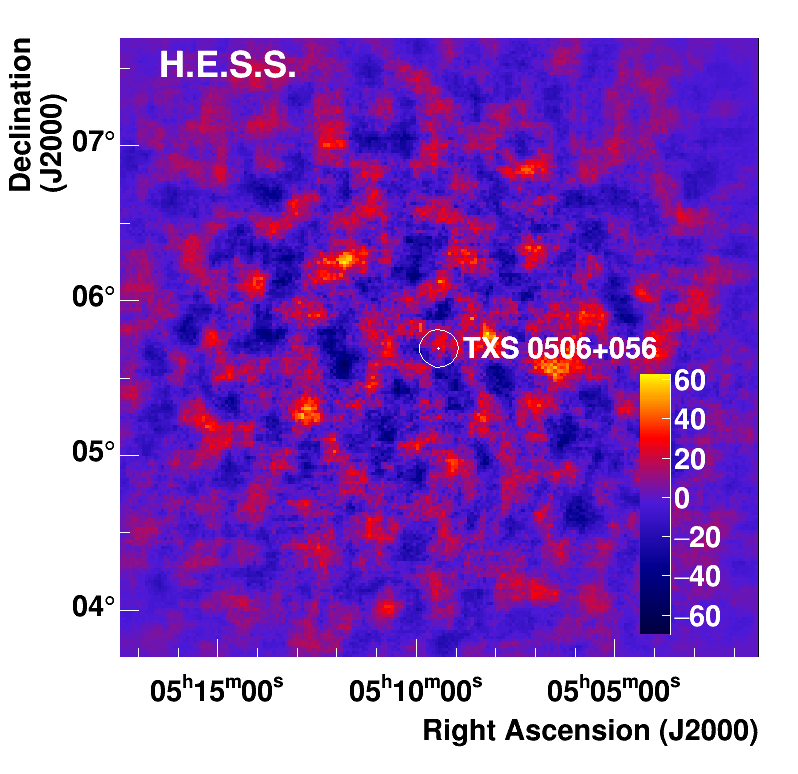}}%
  \quad%
  \subfloat[significance map]{\label{fig:TXSearly:signi}%
    \includegraphics[width=0.48\textwidth]{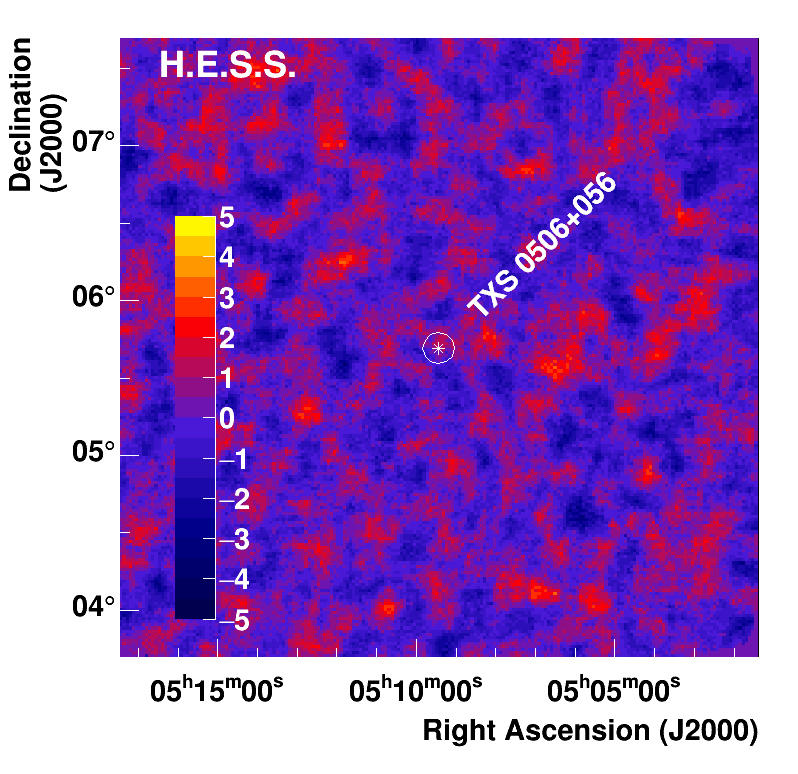}}%
  \caption[IceCube-170922A]{Observations of TXS~0506+056 with the 28m H.E.S.S.-2 telescope obtained during the follow-up of IceCube-170922A in October 2017. No significant signal has been detected in the monoscopic analysis of the 13h dataset.} 
  \label{fig:TXSearly}
\end{figure}
\noindent

Further observations were obtained by H.E.S.S. starting October 12, 2017 once the source became observable again and lasted until October 24, 2017. A total of ~13h of high quality data including the 28m H.E.S.S.-II telescope could be obtained during this period. They were analyzed in monoscopic mode using the Model Analysis~\cite{ModelAnalysis} with {\it loose} cuts to achieve an energy threshold as low as 150~GeV. As shown in Fig.~\ref{fig:TXSearly}, no significant level of gamma-ray emission was detected and upper limits on the VHE gamma-ray flux have been calculated. The spectral assumption for this calculation follow the best fit spectral index of -3.9 as measured by the MAGIC collaboration~\cite{MAGIC170922}. Using a systematic uncertainty of 30\%, limits at 95\% confidence level were derived using {\it TRolke}~\cite{Rolke}. The limits were calculated for each night of the data set individually and are shown in Fig.~\ref{fig:IACTs170922}. All results have been cross-checked with an independent calibration and analysis chain~\cite{ImPACT}, which showed consistent results. 

Taking the obtained limits together with the earlier results published in~\cite{IC-TXS-MMA} and observations from the other IACT collaborations (cf. Fig.~\ref{fig:IACTs170922}) further highlights the rapid variability of the TeV emission of TXS~0506+056.

\begin{figure}[!t]
  \centering
    \includegraphics[width=0.85\textwidth]{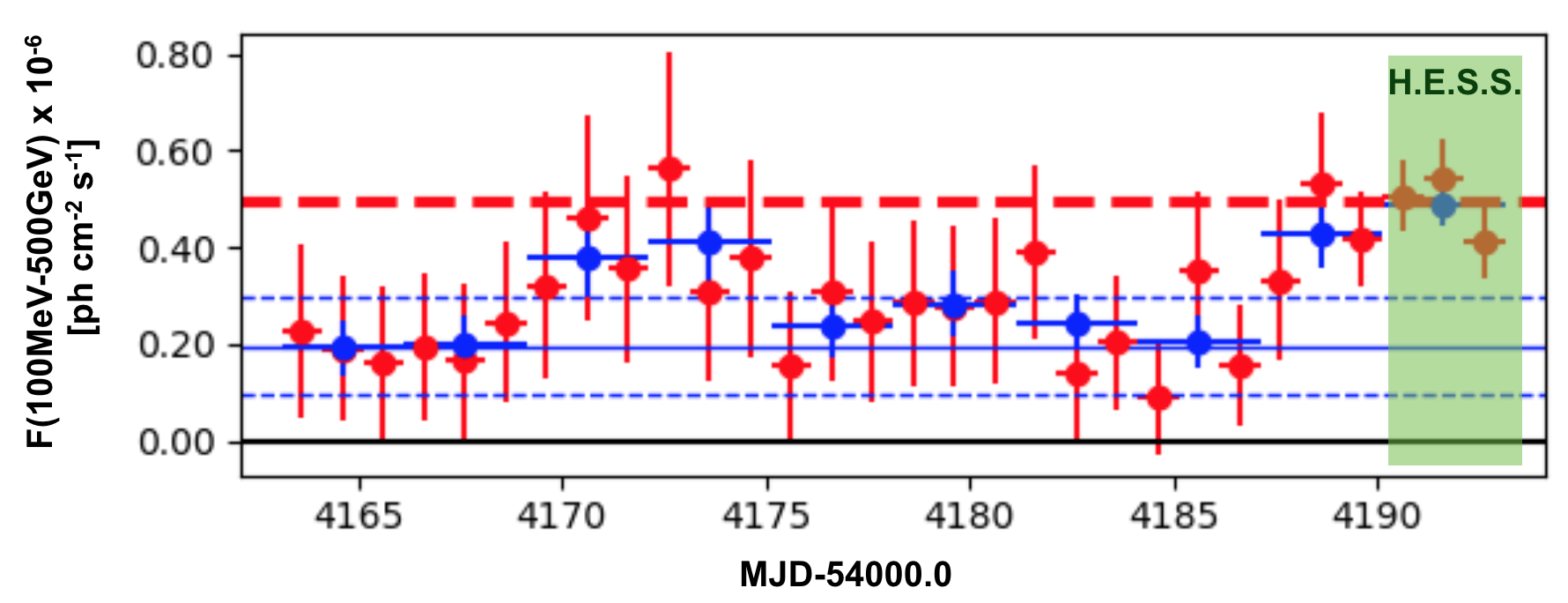}%
  \caption[TXS~0506+056 FLaapLUC]{Fermi-LAT aperture photometry lightcurve of TXS~0506+056 showing the flaring episode in March 2018. The blue solid line denotes the long-term average flux and its statistical uncertainty (blue dashed line). The red, dashed line illustrates 3sigma above the average flux, a criteria to trigger used to trigger the H.E.S.S. ToO observations denoted by the green area. Figure derived using FLaapLUC~\cite{2018A&C....22....9L}.} 
  \label{fig:FLaapLUC}
\end{figure}
\noindent

\subsection{H.E.S.S. observation of TXS~0506+056: follow-up of a Fermi-LAT flare in March 2018}
After the major flaring period of TXS~0506+056 in summer/autumn 2017, two minor flares in the GeV domain have been detected by Fermi-LAT in march 2018. These have been shown up as relative flux variations using FLaapLUC~\cite{2018A&C....22....9L}, an aperture photometry tool built on top of the Fermi Science Tools. The obtained lightcurve in the 100~MeV to 500~GeV band is shown in Fig.~\ref{fig:FLaapLUC}. 

\begin{figure}[!t]
  \centering
   \subfloat[excess map]{\label{fig:TXSMarch2018:excess}
    \includegraphics[width=0.48\textwidth]{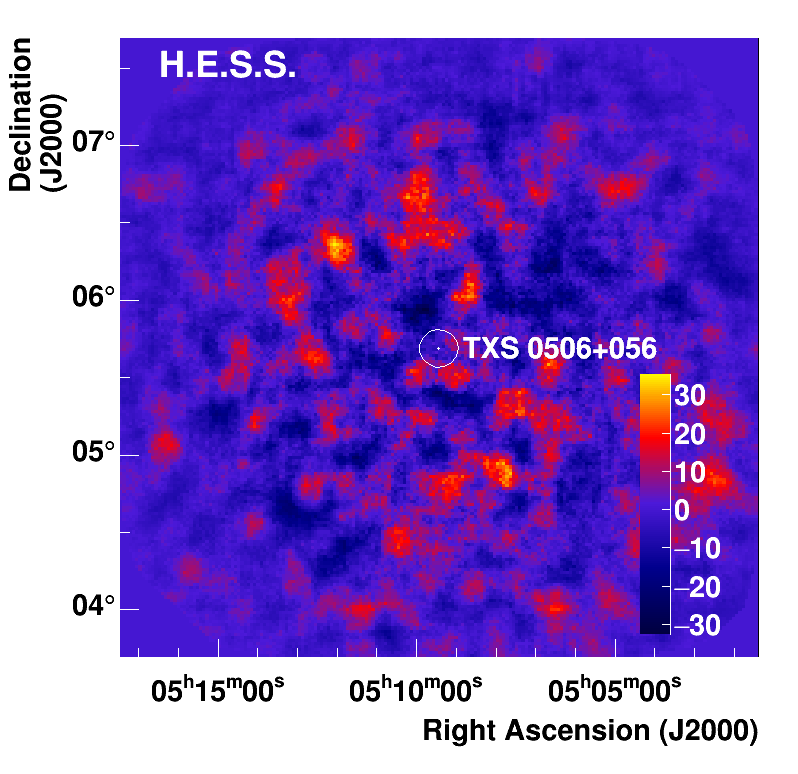}}%
  \quad%
  \subfloat[significance map]{\label{fig:TXSMarch2018:signi}%
    \includegraphics[width=0.48\textwidth]{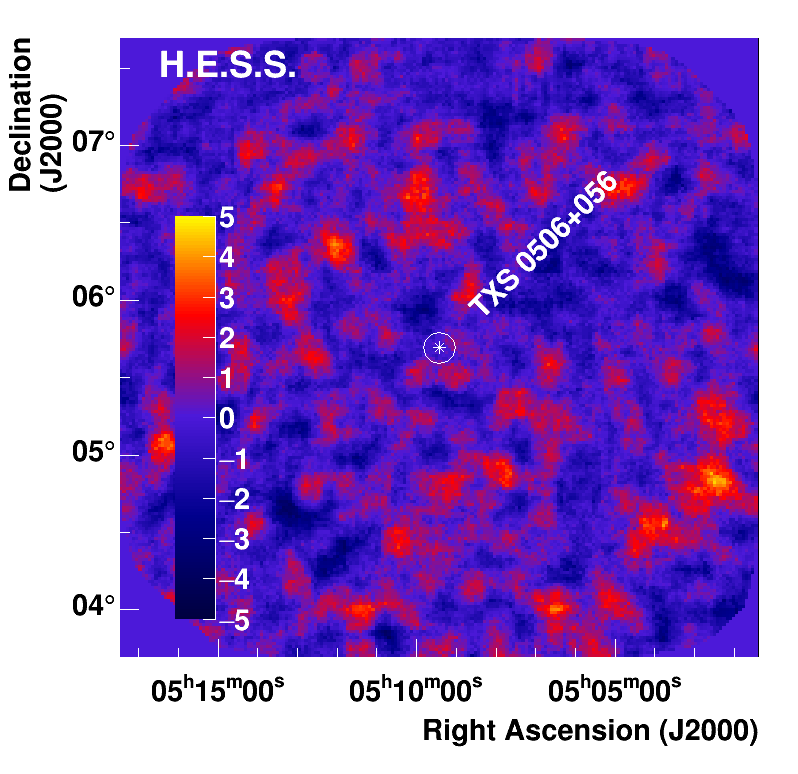}}%
  \caption[TXS~0506+056 March 2018]{Observations of TXS~0506+056 with the 28m H.E.S.S.-2 telescope obtained during the March 2018 flaring period. No significant signal has been detected in the analysis of the 2.4h dataset.} 
  \label{fig:TXSMarch2018}
\end{figure}

While the source could not be observed during the first flaring episode around MJD 58173, ToO observations within the H.E.S.S. AGN ToO program (cf.~\cite{2017ICRC...35..652S}) were triggered on March 14, 2018 (MJD 58191). 2.7h of high quality data could be obtained during the three following nights until March 17, 2018. Again, analyzing the data set in monoscopic mode, reaching an energy threshold of about 140~GeV, no significant gamma-ray emission has been found. The obtained excess and significance maps are shown in Fig.~\ref{fig:TXSMarch2018}. No neutrino excess has been reported by any of the high-energy neutrino telescopes during the period discussed here. Differential upper limits on the VHE flux derived from the H.E.S.S. observations are given in Table~\ref{tab:UL}. They have been derived assuming a spectrum following $E^{-3.9}$ as measured by the MAGIC collaboration during the campaign in September 2018~\cite{MAGIC170922}.

Further observations across the MWL spectrum and especially in the VHE gamma-ray band will allow to characterize the behaviour of TXS~0506+056 in more detail. Of particular interest here is the flaring duty-cycle which influences the calculation of the significance of the correlation between the high-energy neutrino IceCube-170922A and the flaring episode in 2017.

\noindent

\begin{table}[!th]
    \centering
    \begin{tabular}{|c|c|}
    \hline
    energy range [TeV] & flux upper limit [ph/TeV/m${}^{2}$/s] \\ \hline
    0.14 - 0.31     &  2.7e-06\\
    0.31 - 0.66   & 5.9e-08 \\
    0.66 - 1.42 & 3.5e-09 \\
    1.42 - 3.05 & 3.1e-09 \\
    3.05 - 6.55 & 1.1e-09 \\ \hline
    \end{tabular}
    \caption{Upper limits on the VHE gamma-ray flux derived from H.E.S.S. observations of TXS0506+056 in March 2018.}
    \label{tab:UL}
\end{table}

\section{H.E.S.S. and the IceCube Gamma-ray Follow-up program}
Complementing the publicly available information on the detection of single, high-energy neutrinos\footnote{\url{https://gcn.gsfc.nasa.gov/amon.html}}, the online alert system of IceCube (cf.~\cite{2017APh....92...30A}) is also searching for the arrival of temporally correlated neutrino events from a priori defined candidate sources. This analysis pipeline has for example enabled the detection of the neutrino excess from TXS~0506+056 in 2014/2015~\cite{2018Sci...361..147I}. The most significant of these {\it neutrino flares} are used to trigger follow-up observations with VHE gamma-ray instruments in the Gamma-ray Follow-Up (GFU) program~\cite{2016JInst..1111009I}. This cooperation between IceCube and the IACT collaborations has been implemented in 2012 using a neutrino event selection dedicated to the Northern hemisphere. Recently the neutrino selection has been extended to include events from the Southern hemisphere, thus enabling a participation of the H.E.S.S. collaboration in the GFU program. Following the spirit of the GFU framework a dedicated selection of potential joint neutrino and gamma-ray sources has been compiled based on the following criteria:

\begin{itemize}
    \item presence in 3FGL or 3FHL catalog
    \item culmination at HESS site surpassing 30 deg
    \item extragalactic source with known redshift and $z\leq 1.0$
    \item source flagged as {\it variable} in the Fermi-LAT catalogs
    \begin{itemize}
        \item 3FGL: variability > 77.2
        \item 3FHL: variability Bayesian blocks > 1
    \end{itemize}
    \item Allowing for gamma-ray flares with a 10 fold increase over the average Fermi-LAT flux, the extrapolated flux above 100~GeV has to be higher than the H.E.S.S. $5\sigma$ sensitivity for 5h of observations 
\end{itemize}

Complementing this Fermi-LAT based source selection, all H.E.S.S. detected extragalactic sources, Sgr A*, and the Crab nebula have been added to the source list. The final list comprises 139 sources and has been implemented in the IceCube real-time analysis framework end of 2018. 

The reaction to IceCube alerts on neutrino flares from these sources has been implemented and commissioned in the fully automatic H.E.S.S. alert system with the beginning of 2019. First alerts have been received and follow-up observations have been obtained. The analysis is in progress and results will be published elsewhere.

\section{Acknowledgements}
The support of the Namibian authorities and of the University of Namibia in facilitating the construction and operation of H.E.S.S. is gratefully acknowledged, as is the support by the German Ministry for Education and Research (BMBF), the Max Planck Society, the German Research Foundation (DFG), the Helmholtz Association, the Alexander von Humboldt Foundation, the French Ministry of Higher Education, Research and Innovation, the Centre National de la Recherche Scientifique (CNRS/IN2P3 and CNRS/INSU), the Commissariat \`a l'\'energie atomique et aux \'energies alternatives (CEA), the U.K. Science and Technology Facilities Council (STFC), the Knut and Alice Wallenberg Foundation, the National Science Centre, Poland grant no. 2016/22/ M/ST9/00382, the South African Department of Science and Technology and National Research Foundation, the University of Namibia, the National Commission on Research, Science \& Technology of Namibia (NCRST), the Austrian Federal Ministry of Education, Science and Research and the Austrian Science Fund (FWF), the Australian Research Council (ARC), the Japan Society for the Promotion of Science and by the University of Amsterdam. We appreciate the excellent work of the technical support staff in Berlin, Zeuthen, Heidelberg, Palaiseau, Paris, Saclay, T\"ubingen and in Namibia in the construction and operation of the equipment. This work benefited from services provided by the H.E.S.S. Virtual Organisation, supported by the national resource providers of the EGI Federation.

\bibliographystyle{JHEP}
\bibliography{ref.bib}

\end{document}